\documentclass[pra,aps,10pt,superscriptaddress,twocolumn,floatfix]{revtex4-1}
\usepackage{graphicx,color}
\usepackage{amsmath,amssymb,bm}
\usepackage{simplewick}
\usepackage{braket}		% Dirac notation

\usepackage[plainpages=false,pdfpagelabels,colorlinks=true,linkcolor=red,urlcolor=blue,citecolor=blue,pdftitle={Entanglement Entropy of Eigenstates of Quadratic Fermionic Hamiltonians},pdfauthor={},pdfdisplaydoctitle=true,pdfduplex=DuplexFlipLongEdge]{hyperref}

\definecolor{darkred}{rgb}{0.90,0.2,0.2}
\definecolor{darkgreen}{rgb}{0,0.60,.2}
\definecolor{darkblue}{rgb}{0.1,0.3,1}
\definecolor{grey}{cmyk}{0,0,0,0.25}
\definecolor{orange}{cmyk}{0,0.6,0.8,0}

\begin{document}
\title{Entanglement Entropy of Eigenstates of Quadratic Fermionic Hamiltonians}

\author{Lev Vidmar}
\affiliation{Department of Physics, The Pennsylvania State University, University Park, PA 16802, USA}
\author{Lucas Hackl}
\affiliation{Department of Physics, The Pennsylvania State University, University Park, PA 16802, USA}
\affiliation{Institute for Gravitation and the Cosmos, The Pennsylvania State University, University Park, PA 16802, USA}
\author{Eugenio Bianchi}
\affiliation{Department of Physics, The Pennsylvania State University, University Park, PA 16802, USA}
\affiliation{Institute for Gravitation and the Cosmos, The Pennsylvania State University, University Park, PA 16802, USA}
\author{Marcos Rigol}
\affiliation{Department of Physics, The Pennsylvania State University, University Park, PA 16802, USA}

\begin{abstract}
In a seminal paper [D. N. Page, \href{https://doi.org/10.1103/PhysRevLett.71.1291}{Phys.~Rev.~Lett.~{\bf 71},~1291~(1993)}], Page proved that the average entanglement entropy of subsystems of random pure states is $S_{\rm ave}\simeq\ln{\cal D}_{\rm A} - (1/2) {\cal D}_{\rm A}^2/{\cal D}$ for $1\ll{\cal D}_{\rm A}\leq\sqrt{\cal D}$, where ${\cal D}_{\rm A}$ and ${\cal D}$ are the Hilbert space dimensions of the subsystem and the system, respectively. Hence, typical pure states are (nearly) maximally entangled. We develop tools to compute the average entanglement entropy $\langle S\rangle$ of all eigenstates of quadratic fermionic Hamiltonians. In particular, we derive exact bounds for the most general translationally invariant models $\ln{\cal D}_{\rm A} - (\ln{\cal D}_{\rm A})^2/\ln{\cal D} \leq \langle S \rangle \leq \ln{\cal D}_{\rm A} - [1/(2\ln2)] (\ln{\cal D}_{\rm A})^2/\ln{\cal D}$. Consequently we prove that: (i) if the subsystem size is a finite fraction of the system size then $\langle S\rangle<\ln{\cal D}_{\rm A}$ in the thermodynamic limit, i.e., the average over eigenstates of the Hamiltonian departs from the result for typical pure states, and (ii) in the limit in which the subsystem size is a vanishing fraction of the system size, the average entanglement entropy is maximal, i.e., typical eigenstates of such Hamiltonians exhibit eigenstate thermalization.
\end{abstract}

\maketitle

\begin{figure}[!bt]
\begin{center}
\includegraphics[width=0.99\columnwidth]{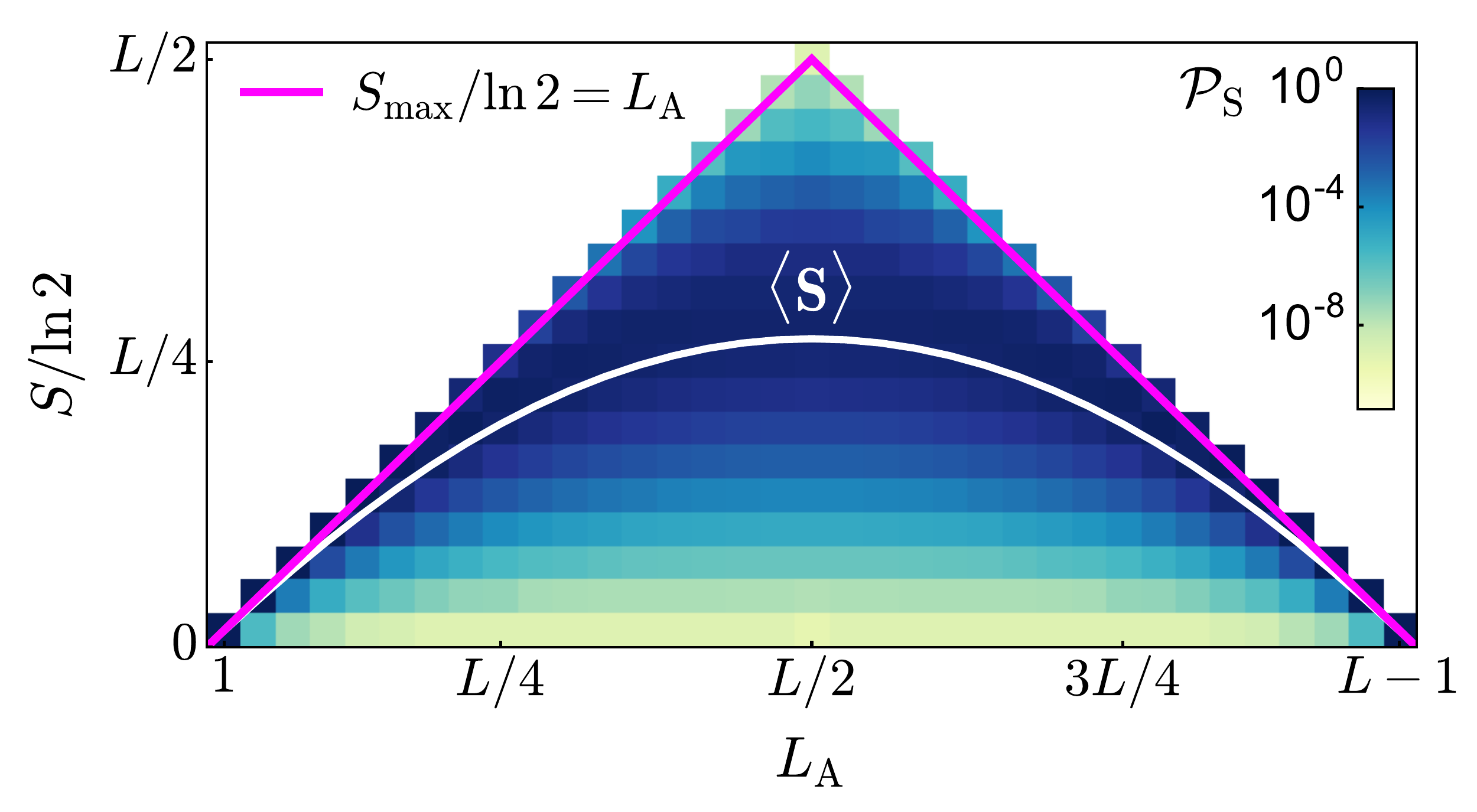}
\vspace*{-0.5cm}
\caption{
{\it Entanglement entropy of eigenstates of noninteracting spinless fermions in a periodic chain with $L=36$ sites.}
Results are plotted as a function of the linear subsystem size $L_{\rm A}$. (Lower line) Average entanglement entropy $\langle S\rangle$ of all eigenstates, and (upper line) upper bound  $S_{\rm max} = \ln{\cal D}_{\rm A} = L_{\rm A} \ln2$ for $L_{\rm A} \leq L/2$ [$S_{\rm max}=(L-L_{\rm A})\ln 2$ for $L_{\rm A} > L/2$]. Each pixel denotes the weight of eigenstates $|m\rangle$ with target entropy $S$, defined as ${\cal P}_S = {\cal D}^{-1} \sum_{0 \leq (S - S_m) < \ln 2}$.
} 
\label{fig1}
\end{center}
\end{figure}

\emph{Introduction}.
The concept of entanglement is a cornerstone in modern quantum physics. Different measures of entanglement have been extensively used to probe the structure of pure quantum states~\cite{eisert_cramer_10}, and they have started to be measured in experiments with ultracold atoms in optical lattices~\cite{islam_ma_15, kaufman_tai_16}. Here, we are interested in the bipartite entanglement entropy (referred to as the entanglement entropy) in fermionic lattice systems. In such systems, an upper bound for the entanglement entropy of a subsystem A (smaller or equal than its complement) is $S_{\rm max} = \ln{\cal D}_{\rm A}$, where ${\cal D}$ and ${\cal D}_{\rm A}$ (${\cal D}_{\rm A}\leq\sqrt{\cal D}$) are the dimensions of the Hilbert space of the system and of the subsystem (see Fig.~\ref{fig1} for an example for spinless fermions). Note that $\ln{\cal D}_{\rm A}\propto V_{\rm A}$, where $V_{\rm A}$ is the number of sites in A, i.e., this upper bound scales with the ``volume'' of A. (When A is larger than its complement, the Hilbert space of the complement is the one that determines $S$.) Almost 24 years ago, motivated by the puzzle of information in black hole radiation \cite{page93a}, Page proved~\cite{page93} that typical (with respect to the Haar measure) pure states nearly saturate that bound~\cite{lubkin78, lloyd_pagels_88, foong_kanno_94, sanchezruiz95, sen96}. Their reduced density matrices are thermal at infinite temperature~\cite{goldstein_lebowitz_06, popescu_short_06, tasaki_98}.

In stark contrast with typical pure states, ground states and low-lying excited states of local Hamiltonians are known to exhibit an {\it area-law} entanglement~\cite{eisert_cramer_10}. Namely, their entanglement entropy scales with the area of the boundary of the subsystem. On the other hand, most eigenstates of local Hamiltonians at nonzero energy densities above the ground state are expected to have a volume-law entanglement entropy (with the exception of many-body localized systems \cite{nandkishore_huse_review_15, altman_vosk_review_15}). Within the eigenstate thermalization hypothesis (ETH) \cite{deutsch_91, srednicki_94, rigol_dunjko_08}, one expects volume-law entanglement in all eigenstates (excluding those at the edges of the spectrum) of quantum chaotic Hamiltonians \cite{santos_polkovnikov_12, deutsch13, garrison15, beugeling15, dalessio_kafri_16}, with those in the center of the spectrum exhibiting maximal entanglement~\cite{dalessio_kafri_16}.

Thanks to the availability of powerful analytical and computational tools to study ground states, many remarkable results have been obtained for the entanglement entropy of such states \cite{srednicki_93, osterloh_amico_2002, osborne_nielsen_02, vidal_latorre_03, calabrese_cardy_04, hastings_07}. On the other hand, for excited states there is a wide gap between what is expected and what has been shown. For interacting Hamiltonians, computational studies are severely limited by finite-size effects so it is difficult to know what happens to the entanglement entropy with increasing the subsystem size. This question was recently addressed for quadratic~\cite{storms14, lai15, nandy_sen_16} and nonquadratic but integrable~\cite{alba15} Hamiltonians, for which one can study much larger lattices, revealing that randomly generated eigenstates are generally maximally entangled in the limit in which the size of the subsystem is a vanishing fraction of the size of the system (in short, a vanishing subsystem fraction).

In this Letter we prove that, for a nonvanishing subsystem fraction, the average entanglement entropy of all many-body fermionic eigenstates of translationally invariant quadratic Hamiltonians departs from Page's result. Still, the average entanglement entropy exhibits a volume law scaling. In the limit of a vanishing subsystem fraction, we prove that the overwhelming majority of eigenstates are maximally entangled. Our proof stands on calculations of spectrum averages of eigenstate entanglement entropies, which are based on the insight that such averages can be obtained as traces over even powers of a correlation matrix, without the need of calculating its eigenvalues.

{\it Entanglement entropy of energy eigenstates.}
We study the most general quadratic Hamiltonian of spinless fermions: $\hat{H} = - \sum_{i,j=1}^V ( \Delta_{ij}\hat{f}_i^\dagger \hat{f}_j^\dagger + \Delta_{ij}^*\hat{f}_j \hat{f}_{i} + t_{ij}\hat{f}_i^\dagger \hat{f}_{j})$, where $\Delta_{ij} = -\Delta_{ji}$ and $t_{ij} = t_{ji}^*$, and $\hat{f}_i$ is the fermionic annihilation operator at site $i$. A Bogoliubov transformation $\hat f_i=\sum_{l=1}^V(\alpha_{il}\hat{c}_l + \beta_{il}\hat{c}_l^\dagger)$ rotates the Hamiltonian so that it commutes with the quasiparticle number operator $\hat{N}_l = 2\,\hat{c}^\dagger_l\,\hat{c}^{\phantom{\dagger}}_l-1$. Hence, the many-body energy eigenkets $|m\rangle$ satisfy $\hat{N}_l|m\rangle=N_l |m\rangle$ with $N_l=\pm 1$, and we adopt the binary representation $m=1+\sum_{l=1}^V \frac{1+N_l}{2}\, 2^{l-1}$ ($m$ runs from $1$ to ${\cal D} = 2^V$, $V$ is the number of lattice sites).

Correlations of a state $|m\rangle$ are encoded in $V\times V$ one-body correlation matrices. They form a $2V\times2V$ matrix $J$, which is a linear complex structure~\cite{suppmat}
\begin{align}\label{eq:lcs}
{\rm i}J&=\left(\begin{array}{c|c}
\langle m|\hat f_i^\dagger \hat f_j - \hat f_j \hat f_i^\dagger|m\rangle & \langle m|\hat f_i^\dagger \hat f_j^\dagger - \hat f_j^\dagger \hat f_i^\dagger|m\rangle\\ \hline \langle m|\hat f_i \hat f_j - \hat f_j \hat f_i|m\rangle & \langle m| \hat f_i \hat f_j^\dagger - \hat f_j^\dagger \hat f_i|m\rangle
\end{array}\right) .
\end{align}
Since the many-body eigenstates $\{|m\rangle\}$ are Gaussian states, the matrix ${\rm i}J$ fully characterizes them~\cite{peschel_03, peschel_eisler_09, dierckx_fannes_08, hackl_bianchi_17}. Correlations of a subsystem $A$ containing $V_{\rm A}$ sites are encoded in the restricted complex structure $[ {\rm i}J ]_{\rm A}$, the $2V_{\rm A}\times 2V_{\rm A}$ matrix obtained by restricting the matrix ${\rm i}J$ to the entries with $i,j\in A$. The entanglement entropy of subsystem $A$ in the eigenstate $|m\rangle$ can be computed as~\cite{peschel_03, peschel_eisler_09}
\begin{align}\label{def_Salpha}
S_m = - \mathrm{Tr} \left\{ \left(\frac{1\!\!1+[{\rm i}J]_A}{2}\right)\ln \left(\frac{1\!\!1+[{\rm i}J]_A}{2}\right) \right\}.
\end{align}

Expanding Eq.~(\ref{def_Salpha}) in powers of $[{\rm i}J]_{\rm A}$, about $[{\rm i}J]_{\rm A}=0$, allows one to compute the entanglement entropy without calculating the eigenvalues of $[ {\rm i}J ]_{\rm A}$
\begin{equation} \label{def_Salphaexp}
 S_{m} = V_{\rm A} \ln2 - \sum_{n=1}^\infty \frac{{\rm Tr} [{\rm i}J]_{\rm A}^{2n} }{4n(2n-1)},
\end{equation}
where we use the compressed notation ${\rm Tr} [{\rm i}J]_{\rm A}^{2n} \equiv {\rm Tr} \{ [{\rm i}J]_{\rm A}^{2n} \}$. Since the restricted complex structure satisfies $[{\rm i}J]_{\rm A}^2\leq 1\!\!1$~\cite{suppmat}, one has 
\begin{equation}
0\leq \mathrm{Tr} [{\rm i}J]_{\rm A}^{2(m+1)} \leq \mathrm{Tr} [{\rm i}J]_{\rm A}^{2m} \leq 2V_{\rm A}
\label{eq:ineq}
\end{equation}
and the series in Eq.~(\ref{def_Salphaexp}) is convergent.

Equation~(\ref{def_Salphaexp}) allows one to compute the average over the ensemble of all eigenstates $\{|m\rangle\}$ as
\begin{equation} \label{def_Savr}
\langle S\rangle = V_{\rm A} \ln2 - \sum_{n=1}^\infty \frac{\langle{\rm Tr} [{\rm i}J]_{\rm A}^{2n}\rangle }{4n(2n-1)}\,,
\end{equation}
where we define $\langle O \rangle\equiv{\cal D}^{-1} \sum_{m=1}^{\cal D} O_m$.

A remarkable property of the series in Eq.~(\ref{def_Savr}) is that every higher-order term lowers the average entanglement entropy. Hence, any truncation gives an upper bound. Using the inequality in Eq.~(\ref{eq:ineq}), one can also produce lower bounds for the average entanglement entropy. To obtain the first-order lower and upper bounds, $S^-_1 \leq \langle S \rangle \leq S^+_1$, we only need to compute $\langle\mathrm{Tr} [{\rm i}J]_{\rm A}^{2} \rangle$ since: (i) truncating the series in Eq.~(\ref{def_Savr}) after the first term results in $S^+_1$, and (ii) substituting all averages of higher-order traces by $\langle\mathrm{Tr} [{\rm i}J]_{\rm A}^{2} \rangle$ results in $S^-_1$. This gives 
\begin{equation}
V_{\rm A}\ln{2} -\frac{\langle\mathrm{Tr} [{\rm i}J]_{\rm A}^{2} \rangle}{2}\ln{2} \leq \langle S \rangle \leq V_{\rm A}\ln{2}-\frac{\langle\mathrm{Tr} [{\rm i}J]_{\rm A}^{2} \rangle}{4}\,.
\end{equation}

For a given eigenstate $|m\rangle$ of our Hamiltonian, $[{\rm i}J]_{\rm A}$ is linear in the quantum numbers $N_l$. In fact, $\langle m |\hat f_i^\dagger \hat f_j - \hat f_j \hat f_i^\dagger| m\rangle = \sum_{l}N_l(\alpha^*_{il}\alpha_{jl} - \beta_{il}^*\beta_{jl})$ and $\langle m|\hat f_i^\dagger \hat f_j^\dagger - \hat f_j^\dagger \hat f_i^\dagger| m\rangle = \sum_l N_l(\alpha_{il}^*\beta_{jl}^* - \beta_{il}^*\alpha_{jl}^*)$. The average $\langle{\rm Tr} [{\rm i}J]_{\rm A}^{2n}\rangle$ can therefore be computed from the binomial correlation function $\langle N_{l_1}\cdots N_{l_{2n}}\rangle$. In particular, to compute $\langle\mathrm{Tr} [{\rm i}J]_{\rm A}^{2} \rangle$, we use that $\langle N_{l}N_{l'}\rangle=\delta_{ll'}$ to get
\begin{equation}\label{gen_1st}
\langle\mathrm{Tr} [{\rm i}J]_{\rm A}^{2} \rangle=2\sum_{l=1}^V\sum_{i,j\in A}\!\!\begin{array}{l}
\\
\left(|\alpha_{il}^*\alpha_{jl}-\beta_{il}^*\beta_{jl}|^2\right.\\
\,\,\,\left.+|\alpha_{il}\beta_{jl}-\beta_{il}\alpha_{jl}|^2\right).
\end{array}
\end{equation}
Whenever $\langle\mathrm{Tr} [{\rm i}J]_{\rm A}^{2} \rangle/V_{\rm A}$ does not vanish in the thermodynamic limit, $\langle S \rangle/ V_{\rm A}<\ln{2}$.

{\it Bounds for translationally invariant Hamiltonians.}
The Bogoliubov coefficients for a translationally invariant system in $d$ dimensions are: $\alpha_{il} = e^{{\rm i}\vec{k}_l\cdot\vec{x}_i}u_{\vec{k}_l}/\sqrt{V}$ and $\beta_{il}=e^{-{\rm i}\vec{k}_l\cdot\vec{x}_i}v_{\vec{k}_l}/\sqrt{V}$, with $u_{\vec{k}}=u_{-\vec{k}}$, $v_{\vec{k}}=-v_{-\vec{k}}$, and $|u_{\vec{k}}|^2+|v_{\vec{k}}|^2=1$. Substituting these in Eq. (\ref{gen_1st}) leads to
\begin{align}
\langle \mathrm{Tr}[iJ]^2_{\rm A}\rangle&=\frac{2V_{\rm A}^2}{V}-\sum_{l=1}^V\sum_{i,j\in {\rm A}}\frac{8|v_{\vec{k}_l}|^2|u_{\vec{k}_l}|^2\cos{2\vec{k}_l(\vec{x}_i-\vec{x}_j)}}{V^2}\nonumber\\ &=\frac{2V_{\rm A}^2}{V}-\sum_{l=1}^V\frac{8|v_{\vec{k}_l}|^2|u_{\vec{k}_l}|^2}{V^2}\prod_{\mathcal{\eta}=1}^d\frac{\sin^2(L_\eta [k_l]_\eta)}{\sin^2([k_l]_\eta)}\,,
\end{align}
where the space sum runs within an $d$-dimensional hypercube $A$ with side lengths $L_\eta$ containing $V_{\rm A}=\prod_{\eta=1}^d L_\eta$ sites. One can bound $\langle \mathrm{Tr}[{\rm i}J]^2_A\rangle$ from below using $|u_{\vec{k}_l}|^2|v_{\vec{k}_l}|^2 \leq 1/4$. As $V\rightarrow\infty$, one can substitute $\sum_l\to \frac{V}{(2\pi)^d}(\int^\pi_{-\pi}\!\!dk)^d$. Since $\int^\pi_{-\pi}\!\!dk \sin^2(L_\eta k)/\sin^2(k)=2\pi L_\eta$ then $\langle \mathrm{Tr}[{\rm i}J]^2_A\rangle \geq 2V_A^2/V - 2V_A/V$. In the thermodynamic limit, we get that $\langle \mathrm{Tr}[{\rm i}J]^2_A\rangle = 2V_A^2/V$. The corresponding {\it universal} first-order bounds are:
\begin{align} 
S^+_1 &= V_{\rm A} \ln 2 - \frac{1}{2} \frac{V_{\rm A}^2}{V} = \ln{\cal D}_{\rm A} - \frac{1}{2\ln2} \frac{(\ln{\cal D}_{\rm A})^2}{\ln{\cal D}} \, \nonumber\\
S^-_1 &= V_{\rm A} \ln 2 \left( 1- \frac{V_{\rm A}}{V} \right) = \ln{\cal D}_{\rm A} - \frac{(\ln{\cal D}_{\rm A})^2}{\ln{\cal D}}  \, .\label{Savr1}
\end{align}
Note that: (i) $S^+_1$ and $S^-_1$ fulfill a volume law as they are proportional to $V_{\rm A}$, and (ii) for any nonvanishing subsystem fraction, $\lim_{V \to \infty}V_{\rm A}/V \neq 0$, $S^+_1<V_{\rm A} \ln 2$ in the thermodynamic limit, i.e., the average departs from the result for typical states in the Hilbert space. If the subsystem fraction vanishes in the thermodynamic limit, $\lim_{V \to \infty}V_{\rm A}/V = 0$, the lower and the upper bounds coincide and $\lim_{V_{\rm A}/V \to 0} S^-_{1} = \lim_{V_{\rm A}/V \to 0} S^+_{1} = V_{\rm A} \ln2$. In this limit, the average entanglement entropy is maximal, i.e., typical eigenstates of the Hamiltonian have a typical ({\it \`a la} Page~\cite{page93}) entanglement entropy.

{\it Entanglement entropy bounds for free fermions.}
We now apply our construction to free fermions on a translationally invariant chain with $L$ lattice sites, described by the Hamiltonian $\hat{H}=-\sum_{i,j=1}^L t_{j-i}\hat{f}_i^\dagger \hat{f}_{j}$. In this case, the Bogoliubov coefficients are $u_{k}=1$ and $v_{k}=0$, so that the eigenstates are plane waves. This allows us to obtain closed form expressions for finite systems. We denote the linear subsystem size as $L_{\rm A}$. Figure~\ref{fig1} shows the distribution of $S_m$ for all eigenstates $|m\rangle$ in a lattice with $L=36$ sites, as well as the corresponding average $\langle S \rangle$. It is remarkable that when $L_{\rm A}$ departs from 1, the entanglement entropy of the eigenstates with the most weight departs from $S_{\rm max} = L_{\rm A}\ln 2$~\cite{storms14, nandy_sen_16}.

Using $\langle N_kN_{k'}\rangle=\delta_{kk'}$, we can explicitly compute
\begin{equation}
\langle\mathrm{Tr} [{\rm i}J]_{\rm A}^{2} \rangle=\frac{2}{L^2}\!\!\sum_{i,j=1}^{L_{\rm A}} \sum_{k,k'=1}^L \!\langle N_{k}N_{k'}\rangle\, e^{{\rm i}\frac{2\pi}{L}(k-k')(i-j)}=\frac{2L_{\rm A}^2}{L}
\label{trace2}
\end{equation}
for finite systems. The first-order bounds for $\langle S \rangle$, for finite systems, are then given by Eq.~(\ref{Savr1}) upon replacing $V\to L$ and $V_{\rm A} \to L_{\rm A}$.

\begin{figure}[!bt]
\begin{center}
\includegraphics[width=0.99\columnwidth]{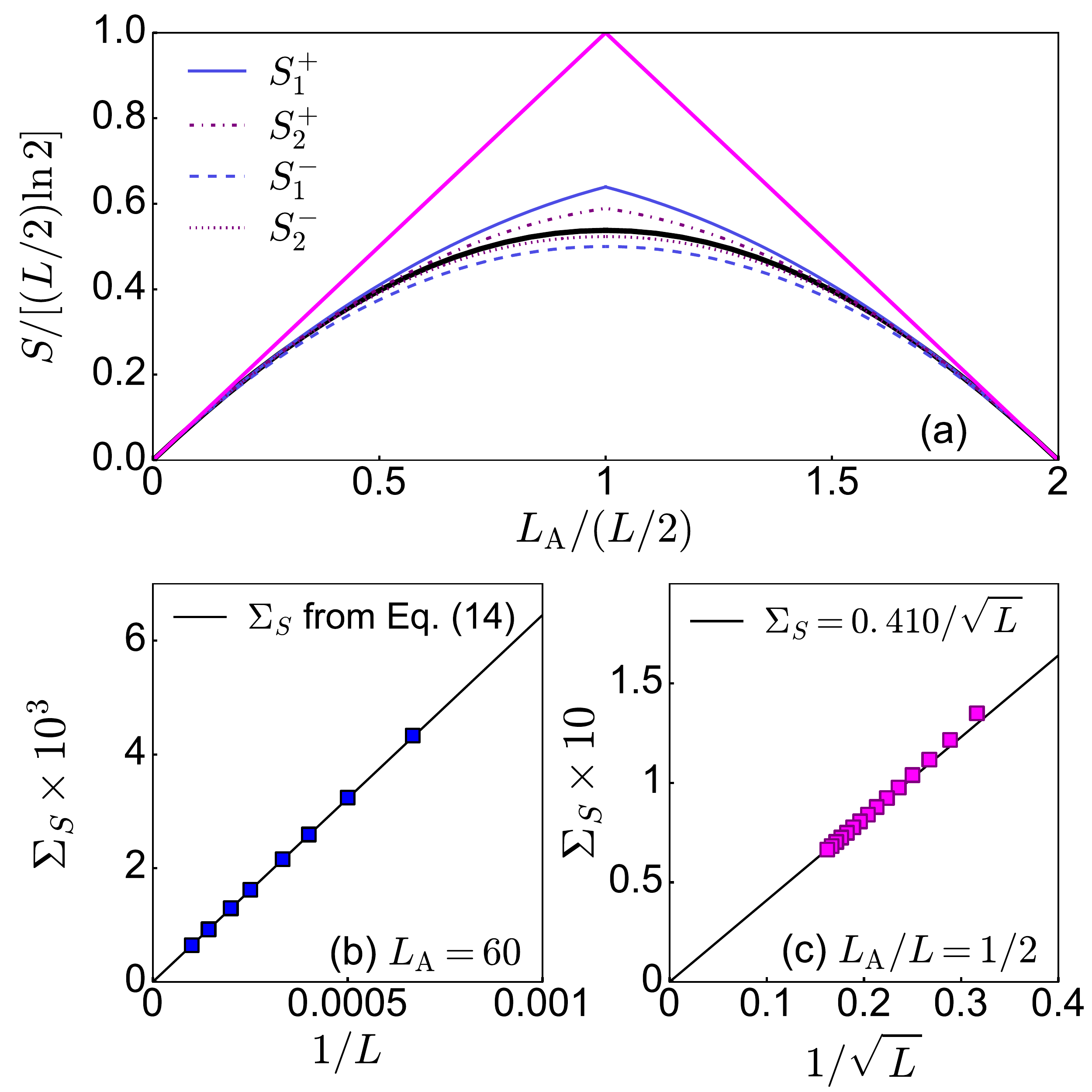}
\vspace*{-0.5cm}
\caption{
{\it Entanglement entropy mean, bounds, and variance for free fermions on a periodic chain.}
(a) Upper bounds ($S^+_{1}$, $S^+_{2}$) and lower bounds ($S^-_{1}$, $S^-_{2}$), given by Eqs.~(\ref{Savr1}) and~(\ref{Savr2}), for the average entanglement entropy. The upper (magenta) line is the maximal entanglement entropy $S_{\rm max}$ and the thick black line is the average entanglement entropy $\langle S \rangle$ on a lattice with $L=36$ sites (same results as in Fig.~\ref{fig1}). (b) $\Sigma_S$ for $L_{\rm A} = 60$ in ensembles of $10^6$ randomly sampled eigenstates and, solid line, the prediction from Eq.~(\ref{SigmaS}). (c) $\Sigma_S$ for $L_{\rm A}/L = 1/2$ calculated using all eigenstates in lattices with $L\leq 38$. The solid line is a single-parameter fit to $\Sigma_S = a/\sqrt{L}$ for $L\geq 30$, with $a =0.410$.
} 
\label{fig2}
\end{center}
\end{figure}

It is straightforward to calculate bounds beyond the first order. A general procedure to compute averages of traces of $[{\rm i}J]_{\rm A}^{2n}$ is presented in Ref.~\cite{suppmat}. The main insight from our analysis is that the term $\langle {\rm Tr} [{\rm i}J]_{\rm A}^{2n} \rangle/L_{\rm A}$ is a polynomial that, when $L \to \infty$, only contains powers from $(L_{\rm A}/L)^n$ to $(L_{\rm A}/L)^{2n-1}$. For the second order upper ($S^+_{2}$) and lower ($S^-_{2}$) bounds~\cite{suppmat}, one gets
\begin{eqnarray} \label{Savr2}
 S^+_{2} & = & L_{\rm A} \ln 2 - \frac{1}{2} \frac{L_{\rm A}^2}{L} - \frac{2}{9} \frac{L_{\rm A}^3}{L^2} + \frac{1}{6} \frac{L_{\rm A}^4}{L^3},  \\[1em]
 S^-_{2} & = & L_{\rm A} \ln 2  - \frac{1}{2} \frac{L_{\rm A}^2}{L} - (2\ln2 - 1) \left( \frac{4}{3} \frac{L_{\rm A}^3}{L^2} - \frac{L_{\rm A}^4}{L^3} \right) \,. \nonumber
\end{eqnarray}
In order to obtain Eq.~\eqref{Savr2}, we neglected finite-size corrections of order ${\cal O}(1/L)$ and higher.

In Fig.~\ref{fig2}(a), we compare the first- and second-order bounds with the average $\langle S \rangle$ computed on a lattice with $L=36$ sites. The bounds can be seen to be very close to the numerically computed average. At $L_{\rm A}/L = 1/2$, where the relative deviation is largest, we get that $0.52 < \langle S \rangle /[(L/2)\ln 2] < 0.59$. In Fig.~S1 of Ref.~\cite{suppmat}, we extrapolate numerical results for $\langle S \rangle$ to the limit $L \to \infty$. Finite-size effects are found to be exponentially small in $L$ \cite{iyer15}, and we obtain $\lim_{L \to \infty} \langle S \rangle/[(L/2) \ln2] = 0.5378(1)$.

{\it Entanglement entropy variance for free fermions.}
In order to understand whether the average of the entanglement entropy over all eigenstates is representative of the entanglement entropy of typical eigenstates, we calculate the variance
\begin{align}\label{eq:gen_variance}
\Sigma_S^2= \frac{\langle S^2\rangle-\langle S\rangle^2}{(L_{\rm A} \ln 2)^2} = \frac{1}{(L_{\rm A}\ln 2)^2}\sum_{m,n=1}^\infty\!\! F_{m,n}\,,
\end{align}
where
\begin{equation} \label{def_Fmn}
 F_{m,n} = \frac{\langle\mathrm{Tr} [{\rm i}J]_{\rm A}^{2m} \,\mathrm{Tr} [{\rm i}J]_{\rm A}^{2n} \rangle-\langle\mathrm{Tr} [{\rm i}J]_{\rm A}^{2m} \rangle\,\langle\mathrm{Tr} [{\rm i}J]_{\rm A}^{2n} \rangle}{4m(2m-1)\,4n(2n-1)} \, .
\end{equation}
The computation of $F_{m,n}$ is, in general, a daunting task. However, by using a summation technique to compute higher-order traces~\cite{suppmat}, we are able to extract key properties of $\Sigma_S$. In particular, we are able to prove that $\Sigma_S$ vanishes with increasing the system size as $\Sigma_S \sim 1/\sqrt{L}$ or faster~\cite{suppmat}. Furthermore, in the limit of vanishing subsystem fraction (fixed $L_{\rm A}$ for $L\rightarrow\infty$), we obtain the lowest order term in $L$ to be
\begin{equation} \label{SigmaS}
 \Sigma_S^2 = \frac{1}{L^2} \frac{1}{(\ln2)^2} \left(\frac{L_{\rm A}}{3}+\frac{1}{6L_{\rm A}}\right)\,.
\end{equation}
Numerical results for $\Sigma_S$ in this limit, reported in Fig.~\ref{fig2}(b), confirm the accuracy of this prediction. Numerical results for $L_{\rm A}/L=1/2$, reported in Fig.~\ref{fig2}(c), confirm that $\Sigma_S\sim 1/\sqrt{L}$ for a nonvanishing subsystem fraction. The vanishing of the variance proves that the average and typical entanglement entropies are identical.

\begin{figure}[!bt]
\begin{center}
\includegraphics[width=0.95\columnwidth]{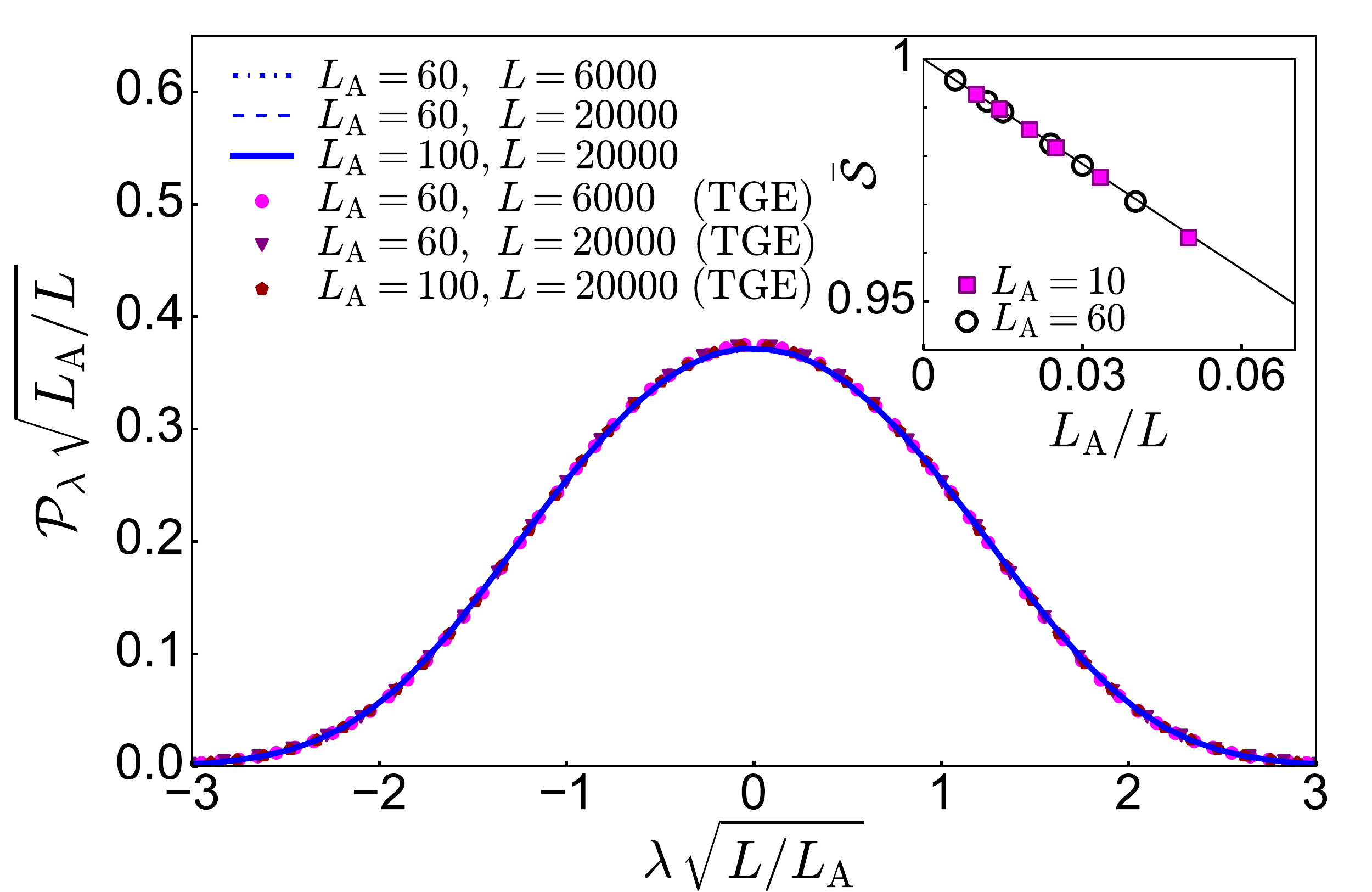}
\caption{
{\it Distribution of eigenvalues of the restricted complex structure for vanishing subsystem fraction.}
The overlapping solid lines depict ${\cal P}_\lambda$, which are averages of the discrete distribution $p_\lambda = \sum_{|\lambda_j -\lambda| < \delta \lambda/2}$ over $10^6$ random eigenstates, where $\lambda_j$ are eigenvalues of $[{\rm i}J]_{\rm A}$ and we take $\delta \lambda = 10^{-2}$. The symbols depict ${\cal P}_\lambda^{\rm (TGE)}$, which are averages of the discrete distribution $p_\lambda^{\rm (TGE)} = \sum_{|\lambda_j -\lambda| < \delta \lambda/2}$ over $10^6$ realizations of the Toeplitz Gaussian ensemble (TGE), where $\lambda_j$ are the eigenvalues of the TGE and we take $\delta \lambda = 10^{-2}$. Both axes are renormalized to show data collapse. (Inset) Scaling of $\bar{\cal S}$ [see Eq.~\eqref{S_vanish2}]. The symbols show numerical results of an average over $10^6$ random eigenstates. The solid line shows the results of Eq.~\eqref{S_vanish2}.
} 
\label{fig3}
\end{center}
\end{figure}

{\it Eigenvalue distribution for free fermions.}
Our results for the average entanglement entropy allow us to unveil some remarkable properties of the eigenvalues $\lambda_j$ of $[{\rm i}J]_{\rm A}$ in energy eigenstates. They satisfy $|\lambda_j|\leq 1$~\cite{suppmat}. It is also straightforward to prove that the average of the sum of eigenvalues vanishes: $\langle \sum_{j} \lambda_j \rangle = \langle {\rm Tr} [{\rm i}J]_{\rm A} \rangle = L_{\rm A}(1 - 2 \langle N \rangle/L) =0 $, where $\langle N \rangle = L/2$ is the average number of particles. On the other hand, the average of the variance of the eigenvalues of $[{\rm i}J]_{\rm A}$ can be calculated using Eq.~(\ref{trace2}), yielding
\begin{equation} \label{def_sigma_lambda}
\sigma^2 = \frac{1}{2L_{\rm A}} \left\langle \sum_j \lambda_j^2 \right\rangle  = \frac{1}{2L_{\rm A}} \langle\mathrm{Tr} [{\rm i}J]_{\rm A}^{2} \rangle = \frac{L_{\rm A}}{L} \, .
\end{equation}
This allows us to conclude that $\sigma^2$ vanishes if $\lim_{L\rightarrow\infty}L_{\rm A}/L = 0$ (implying $\langle S \rangle$ is maximal), and cannot vanish if $\lim_{L\rightarrow\infty}L_{\rm A}/L \neq0$. In Fig.~\ref{fig3}, we report results of numerical calculations of the distribution of eigenvalues of $[{\rm i}J]_{\rm A}$ (for small values of $L_{\rm A}/L$) in a large ensemble of randomly chosen eigenstates. This distribution can be seen to have a universal form that only depends on the ratio $L_{\rm A}/L$, and whose width is $\sqrt{L_{\rm A}/L}$.

The variance of the distribution of eigenvalues of $[{\rm i}J]_{\rm A}$ ($ \sigma^2 = L_{\rm A}/L$) is important as it determines how the average entanglement entropy reaches the maximal value in the thermodynamic as $L_{\rm A}/L\rightarrow0$. The lowest order correction to $\langle S \rangle=L_{\rm A} \ln 2$ in terms of $L_{\rm A}/L$ can be read from Eqs.~\eqref{Savr2}, in which the upper and lower bounds coincide up to ${\cal O}[(L_{\rm A}/L)^2]$,
\begin{equation} \label{S_vanish2}
 \bar{\cal S} \equiv \frac{\langle S \rangle}{L_{\rm A} \ln 2} =  1 - \frac{1}{2\ln2} \frac{L_{\rm A}}{L} + {\cal O} \left[ \left(\frac{L_{\rm A}}{L}\right)^2 \right].
\end{equation}
A comparison of the latter expression to numerical results, reported in the inset in Fig.~\ref{fig3}, yields an almost perfect agreement for $L_{\rm A}/L\lesssim0.05$.

For a vanishingly small subsystem fraction, the facts that (i) the average entanglement entropy is maximal, and (ii) the distribution of eigenvalues of $[{\rm i}J]_{\rm A}$ is universal (see Fig.~\ref{fig3}), hints that a random ensemble may explain those results. We construct such an ensemble, the Toeplitz Gaussian ensemble (TGE). In the TGE, the entries of $[{\rm i}J]_{\rm A}$ are replaced by random complex numbers whose absolute value is that of a normally distributed variable with zero mean and variance $1/L$, and whose phase is uniformly distributed between 0 and $2\pi$. As shown in Fig.~\ref{fig3}, the corresponding eigenvalue distribution is nearly indistinguishable from the numerical calculation over $10^6$ random eigenstates. (See also Fig.~S2 of Ref.~\cite{suppmat}, which shows that taking the limit $L \to \infty$ first, followed by $L_{\rm A} \to \infty$, results in two distributions that are identical.) This shows that, in the limit of vanishing subsystem fraction, no specific information beyond the symmetries of the model appears to be encoded in the restricted complex structure $[{\rm i}J]_{\rm A}$ of typical eigenstates.

{\it Discussion.}
Our work introduces a novel methodology that enables the rigorous study of the entanglement entropy of excited eigenstates of quadratic Hamiltonians. The derivation of exact bounds for the average entanglement entropy of translationally invariant quadratic Hamiltonians reveals a fundamental difference between the results for vanishing and nonvanishing subsystem fractions, which is not captured by the analysis of random pure states in the Hilbert space~\cite{page93}. This highlights the difference in information content between typical eigenstates of physical Hamiltonians, such as those considered here, and typical states in the Hilbert space. The fact that, for a vanishing subsystem fraction, typical eigenstates are maximally entangled constitutes a proof that typical eigenstates satisfy ETH for local observables.

We note that Eqs.~\eqref{eq:lcs}--\eqref{gen_1st} also apply to quadratic models of much current interest such as those appearing in disordered~\cite{khemani_nandkishore_15} and periodically driven (Floquet)~\cite{bukov_dalessio_15} systems. While our study focuses on the von Neumann entanglement entropy, the upper bounds derived remain valid for higher-order Renyi entropies, which are bounded from above by the von Neumann entanglement entropy. This is of particular relevance for current experiments with ultracold atoms on optical lattices~\cite{islam_ma_15, kaufman_tai_16}, which can now measure the second Renyi entropy. 

\medskip

{\it Acknowledgments.} L.V. and M.R. acknowledge support from the Office of Naval Research, Grant No.~N00014-14-1-0540. E.B. acknowledges support from the NSF Grant PHY-1404204. L.H. is supported by a Frymoyer fellowship. This research was supported in part by the Perimeter Institute for Theoretical Physics. The computations were done at the Institute for CyberScience at Penn State.

%%%%%%%%%%%%%%%%%%%%%%%%%%%%%%%%%%%%%%%%   Bibliography

\bibliographystyle{biblev1}
\bibliography{references}

\newpage
\phantom{a}
\newpage
%%%%%%%%%%%%%%%%%%%%%%%%%%%%%%%%%%%%%%%%
\setcounter{figure}{0}
\setcounter{equation}{0}

\renewcommand{\thetable}{S\arabic{table}}
\renewcommand{\thefigure}{S\arabic{figure}}
\renewcommand{\theequation}{S\arabic{equation}}

\renewcommand{\thesection}{S\arabic{section}}

\onecolumngrid

\begin{center}

{\large \bf Supplemental Material:\\
Entanglement Entropy of Eigenstates of Quadratic Fermionic Hamiltonians}\\

\vspace{0.3cm}

Lev Vidmar,$^1$ Lucas Hackl,$^{1,2}$ Eugenio Bianchi$^{1,2}$, Marcos Rigol$^1$\\
$^1${\it Department of Physics, The Pennsylvania State University, University Park, PA 16802, USA}\\
$^2${\it {Institute for Gravitation and the Cosmos, The Pennsylvania State University, University Park, PA 16802, USA}}

\end{center}

\vspace{0.6cm}

\twocolumngrid

\label{pagesupp}

\section{Linear complex structure}
Equation~(1) in the main text introduces the linear complex structure $J$  as a way to parametrize eigenstates of quadratic Hamiltonians. $J$ is best understood as a linear map on the vector space $W_{\mathbb{C}}=\mathbb{C}^{2V}$ spanned by $V$ creation operators $\hat f_i^\dagger$ and $V$ annihilation operators $\hat f_i$. Its eigenvalues are given by $\pm {\rm i}$ with equal multiplicity leading to a decomposition $W_\mathbb{C}=W^+\oplus W^-$ into a direct sum of eigenspaces. The two eigenspaces are related by complex conjugation and define new sets of creation and annihilation operators. This means that every vector in $W^-$ represents a linear combination of a new set of annihilation operators, while vectors in $W^+$ represent linear combinations of the conjugate creation operators. We can use $J$ to write the projector $\frac{1}{2}\left(1\!\!1+{\rm i}J\right)$ that maps $W_{\mathbb{C}}$ onto $W^-$. If we define the vector $\hat{\xi}^a=(\hat f^\dagger_i, \hat f_j)$ consisting of the $V$ creation and $V$ annihilation operators, we can write the condition
\begin{align}
	\frac{1}{2}\sum_{b}\left(1\!\!1+{\rm i}J\right)^{a}{}_b\,\hat{\xi}^b|m\rangle=0\,,
\end{align}
that defines $|m\rangle$ as the state that is annihilated by all operators in $W^-$.

Linear complex structures are used in mathematics to represent the imaginary unit ${\rm i}$ on an even dimensional real vector space. In particular, we have $J^2=-1\!\!1$. Complex structures are used in bosonic and fermionic systems to define Gaussian states. They are ideal for the study of entanglement in subsystems because their restriction to a subsystem encodes all relevant correlations.

To compute the entanglement entropy of a subsystem of size $V_A$, we need to restrict $J$ to the $2V_A\times 2V_A$ matrix $[J]_A$ associated to the degrees of freedom $\hat f_i^\dagger$ and $\hat f_j$ in $A$. Let us prove that the eigenvalues $\lambda_j$ of the restricted matrix $[{\rm i}J]_A$ are in the interval $[-1,1]$. The action of $[{\rm i}J]_A$ onto a column vector $\vec{w}_A=(w_1,\cdots,w_{2V_A})^\intercal$ is
\begin{align}
	[{\rm i}J]_{\rm A}\vec{w}_A=P_A J(w_1,\cdots,w_{2V_A},0,\cdots,0)^\intercal\,,
\end{align}
where $P_A(w_1,\cdots,w_V)^\intercal=(w_1,\cdots,w_{2V_A})^\intercal$ is the orthogonal projection (with respect to the standard inner product) onto the first $V_A$ components. This implies $\lVert [{\rm i}J]_A\vec{w}_A\rVert\leq \lVert\vec{w}\rVert$ because: (i) ${\rm i}J$ is norm-preserving being a symmetric matrix with eigenvalues $\pm 1$, and (ii) the orthogonal projection $P_A$ cannot increase the norm of a vector. This implies that the eigenvalues $\lambda_j$ of $[{\rm i}J]_A$ satisfy $-1\leq \lambda_j\leq 1$.

\section{Traces of even powers of the complex structure for free fermions}
\label{sec_traces}

We now focus on free fermions in a one-dimensional periodic lattice with $L$ sites and replace $V\to L$ and $V_A\to L_A$. In order to compute the higher-order traces
\begin{align} \label{def_trace_general}
\langle\mathrm{Tr} [{\rm i}J]_A^{2n} \rangle=2\!\!\!\!\!\!\!\sum^{L_A}_{i_1,\cdots,i_{2n}=1}\!\!\!\!\!\!\!\!\langle j(i_1-i_2)\cdots j(i_{2n}-i_1)\rangle\,,
\end{align}
with $j(d_i)= \frac{1}{L} \sum^L_{k=1}e^{{\rm i} \frac{2\pi k d_i}{L}} N_k$, we develop a systematic method of computing higher order correlation functions $\langle j(d_1)\cdots j(d_{2n})\rangle$. Technically, these are Fourier transformed correlation functions of the binomial distribution. They can be computed by adopting a strategy analogous to the one used for computing correlation functions of Gaussian distributions:
\begin{enumerate}
\item The building blocks of correlation functions are the so-called $2n$-contractions given by
	\[
	\ \contraction{}{j}{(d_1)j(d_2)\cdots j(d_{2n-1})}{j}
	\contraction{j(d_1)}{j}{(d_2)\cdots }{j}
	j(d_1)j(d_2)\cdots j(d_{2n-1})j(d_{2n})=c_{n}\frac{\delta(\sum^{2n}_{i=1}d_i)}{L^{2n-1}}\,,
	\]
	where $\delta(D)=1$ if $D=0\pmod{L}$ and zero otherwise, i.e., $D = \sum^{2n}_{i=1}d_i$ is restricted to be an integer multiple of $L$. The prefactors $c_{n}$ can be computed systematically as $c_n=L^{2n-1}\langle j(1)^{2n-1}j(1-2n)\rangle$ by using the explicit form $j(d_i)=\frac{1}{L}\sum^{L}_{k=1}e^{{\rm i}\frac{2\pi k d_i}{L}}N_k$. For the results reported in the main text, it is sufficient to compute $c_1=1$ and $c_2=-2$.
	\item Once the $2n$-contractions are known, one can compute a general correlation function as
	\[
	\begin{split}
	\qquad \langle j(d_1)\cdots j(d_{2n})\rangle&=\sum(\text{all possible contractions})\\
	& =\contraction{}{j}{}{j}jj\cdots\contraction{}{j}{}{j}jj+\cdots+\contraction{}{j}{j\cdots j}{j}
	\contraction{j}{j}{\cdots }{j}
	jj\cdots jj\,,
	\end{split}
	\]
	where each contraction consists of a product of different pairings, quadruplings, etc., of the $2n$ $j$'s.
	This is a generalized Wick theorem where we have not only $2$-contractions, but also higher-order $2n$-contractions.
\end{enumerate}
To illustrate this method, let us apply it to the second order correction containing $\langle\mathrm{Tr} [{\rm i}J]_A^{4} \rangle$.

\subsection{Second order term}
We compute the $4$-point correlation function $\langle j(d_1)j(d_2)j(d_3)j(d_4)\rangle$. We find three $2$-contractions and one $4$-contraction:
\begin{align}
\begin{split}
&\langle j(d_1)j(d_2)j(d_3)j(d_4)\rangle=\contraction{}{j}{}{j}
jj\contraction{}{j}{}{j}
jj+\contraction[2ex]{}{j}{jj}{j}\contraction{j}{j}{}{j}
jjjj+\contraction[2ex]{}{j}{j}{j}\contraction{j}{j}{j}{j}
jjjj+\contraction{}{j}{jj}{j}\contraction{j}{j}{}{j}
jjjj\\
&=c_1^2\frac{\delta(d_1+d_2)\delta(d_3+d_4)}{L^2}+c_1^2\frac{\delta(d_1+d_4)\delta(d_2+d_3)}{L^2}\\&
\quad +c_1^2\frac{\delta(d_1+d_3)\delta(d_2+d_4)}{L^2}+c_2\frac{\delta(\sum^4_{i=1}d_i)}{L^3}\,.
\end{split}
\end{align}
Plugging this expression into our sum for the trace gives
\begin{align}
	\langle\mathrm{Tr}[{\rm i}J]_A^{4}\rangle=\frac{8L_A^3+L_A}{3L^2}-\frac{2L_A^4}{L^3}\,,\label{eq:Order2}
\end{align}
where the first term comes from the sum over the $2$-contractions, while the second term comes from the $4$-contraction for which $\delta(\sum^4_{i=1}d_i)=1$ due to $\sum^4_{i=1}d_i=(x_1-x_2)+\cdots+(x_4-x_1)=0$. Note that, in Eq.~(\ref{eq:Order2}), we assume $L_A\leq L/2$ to avoid the complication that $\delta(d_{i_1}+d_{i_2})$ is also nonzero for $d_{i_1}+d_{i_2}=\pm L$. The term $L_A/(3L^2)$ represents a finite size correction of order $1/L^2$. Finite size corrections are expected to appear also at all higher orders, while they are absent at order $n=1$. 

Using the result in Eq.~(\ref{eq:Order2}) and neglecting finite-size corrections of order ${\cal O}(1/L)$ and higher, we arrive to Eqs.~(11) in the main text. For the lower bound, we replace all higher traces by $\langle\mathrm{Tr} [{\rm i}J]_A^{4} \rangle$.

\subsection{Higher order terms}
To compute higher order traces up to order $n$, we need to determine all prefactors up to $c_n$. Here we investigate the scaling in powers of $L_A/L$, which can appear in $\langle\mathrm{Tr} [{\rm i}J]_A^{2n} \rangle/L_A$. A general $2n$-contraction is schematically given by
\begin{align}
\begin{split}
	&\langle j(d_1)\cdots j(d_{2n})\rangle=\contraction{}{j}{}{j}jj\cdots\contraction{}{j}{}{j}jj+\cdots+\contraction{}{j}{j\cdots j}{j}
	\contraction{j}{j}{\cdots }{j}
	jj\cdots jj\\
	&=c_1^n\frac{\delta(d_1+d_2)\cdots\delta(d_{2n_1}+d_{2n})}{L^n}+\cdots+c_n\frac{\delta(\sum^{2n}_{i=1}d_i)}{L^{2n-1}}\,.\label{eq:GenCor}
\end{split}
\end{align}
To compute $\langle \mathrm{Tr} [{\rm i}J]_A^{2n} \rangle/L_A$, we sum over $x_i$ with $1\leq i\leq 2n$, with $\sum^{2n}_{i=1}d_i$ automatically ensured to vanish, which implies that there is one redundant delta in each addend of Eq.~(\ref{eq:GenCor}).

Let us consider a specific contraction $\mathcal{C}$ of the $2n$ correlation function $\langle j(d_1)\cdots j(d_{2n})\rangle$ that consists of $l$ $2n_i$-contractions with $1\leq i\leq l$. This contraction will give $l$ delta functions $\delta(D_i)$ with $D_i$ being specific sums of the $d_i$s according to the chosen contraction:
\begin{align}
	\mathcal{C}=\prod^l_{i=1}c_{n_i}\frac{\delta(D_i)}{L^{2n_i-1}}\,.
\end{align}
This product of deltas appearing in the sum (\ref{def_trace_general}) gives
\begin{align}
\sum^{L_A}_{x_1,\cdots,x_{2n}}\delta(D_1)\cdots\delta(D_l)={\cal O}(L_A^{2n-l+1})\,,
\end{align}
where we use the fact that each of the $l$ deltas reduces the dimension of the sum by one, except the one that is redundant. Therefore, the sum gives a polynomial of degree $2n-l+1$. Applying this result to all contractions, we find
\begin{align}
\sum^{L_A}_{x_1,\cdots,x_{2n}}\langle j(d_1)\cdots j(d_{2n})\rangle=\frac{{\cal O}(L_A^{n+1})}{L^n}+\cdots+\frac{{\cal O}(L_A^{2n})}{L^{2n-1}}\,,
\end{align}
where ${\cal O}(L_A^{n+1})$ refers to a polynomial of degree $n+1$. Therefore we conclude that $\lim_{L\to\infty} \langle\mathrm{Tr} [{\rm i}J]_A^{2n} \rangle /L_A$ is a polynomial function containing only powers from $(L_A/L)^n$ up to $(L_A/L)^{2n-1}$. This result is instrumental in the analysis of the behavior of the entropy for vanishing subsystem fraction $L_A/L \to 0$. It implies that it is sufficient to compute the entropy function up to order $n$ to get the dominating terms up to order $(L_A/L)^n$.

\subsection{Finite-size analysis of the average entanglement entropy for $L_\text{A}=L/2$}

\begin{figure}[!b]
\begin{center}
\includegraphics[width=0.99\columnwidth]{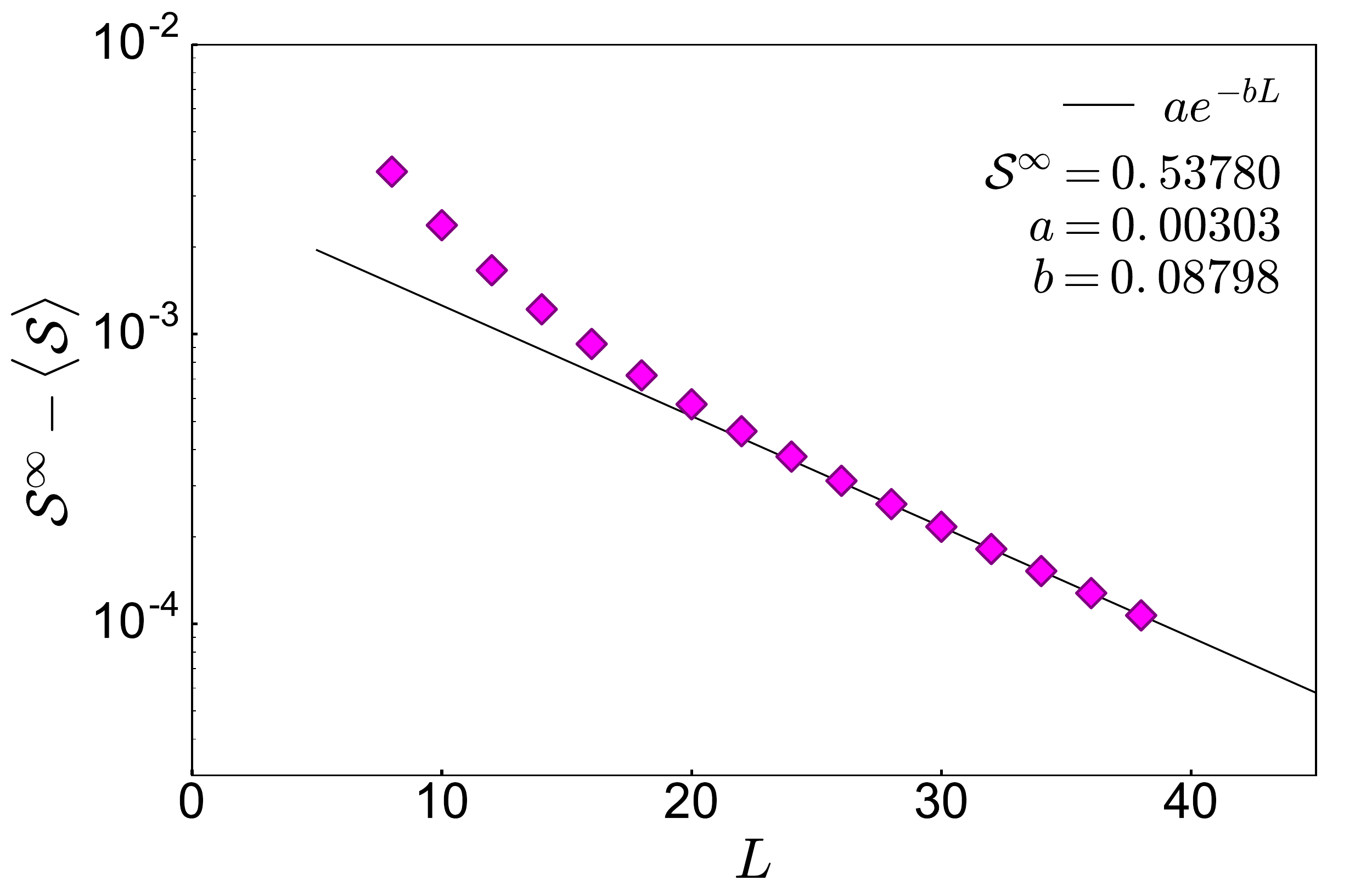}
\caption{{\it Finite-size scaling of the entanglement entropy for $L_A=L/2$.}
Symbols: Subtracted value of the average entanglement entropy $\langle {\cal S}\rangle = \langle S \rangle /[L_A\ln 2]$, i.e., the average over all eigenstates for a given lattice size $L$. The solid line is a three-parameter fit $\langle {\cal S} \rangle = {\cal S}^\infty - a e^{-bL}$ to the data for $L\geq 30$. ${\cal S}^\infty = 0.5378(1)$ is the value of the average entanglement entropy in the thermodynamic limit.
}
\label{figsup1}
\end{center}
\end{figure}

Finite sums of even powers of the traces in Eq.~(\ref{def_trace_general}) provide bounds for the spectrum average of the entanglement entropy $\langle S \rangle$. In Fig.~2(a) of the main text, we compare the bounds in the thermodynamic limit with values of $\langle S \rangle$ in a finite system. Figure~\ref{figsup1}(a) shows the finite-size scaling of $\langle S \rangle$ for $L_A=L/2$. Remarkably, when subtracted by the $L\to\infty$ value ${\cal S}^\infty = \lim_{L\to\infty} \langle S \rangle/[(L/2)\ln 2] = 0.5378(1)$, we see that the finite-size values approach ${\cal S}^\infty$ exponentially fast. This is expected to be the case only in the grand-canonical ensemble in the presence of translational invariance~\cite{iyer15}.

\section{Scaling of the entanglement entropy variance}

Equation~(12) in the main text introduces the entanglement entropy variance $\Sigma_S$, which contains the term
\begin{equation} \label{def_Fmn_2}
 F_{m,n} = \frac{\langle\mathrm{Tr} [{\rm i}J]_A^{2m} \,\mathrm{Tr} [{\rm i}J]_A^{2n} \rangle-\langle\mathrm{Tr} [{\rm i}J]_A^{2m} \rangle\,\langle\mathrm{Tr} [{\rm i}J]_A^{2n} \rangle}{4m(2m-1)\,4n(2n-1)} \, .
\end{equation}
Both terms in the numerator of $F_{m,n}$ can be computed as sums over contractions as discussed in Sec.~\ref{sec_traces}.
While general contractions appear in the first term, the second term contains no contractions between $\mathrm{Tr} [{\rm i}J]_A^{2m} $ and $\mathrm{Tr} [{\rm i}J]_A^{2n}$. We can evaluate the difference by computing all contractions of the first term that contain at least one contraction crossing from the first to the second trace, as all contractions without such a crossing will be canceled by the second term. Let us define the symbol $\langle j(d_1)\cdots j(d_{2m})|j(d_{2m+1})\cdots j(d_{2(m+n)})\rangle$ as the sum over all contractions, for which at least one contraction crosses the separator indicated by the symbol $|$.

The contractions are schematically given by
\begin{align}
\begin{split}
&\langle j\cdots j|j\cdots j jj\cdots jj\rangle\\
&=\contraction{}{j}{\cdots j|}{j}\contraction[2ex]{j\cdots }{j}{|j\cdots }{j}
j\cdots j|j\cdots j\contraction{}{j}{}{j}jj\cdots \contraction{}{j}{}{j}jj+\cdots+\contraction{j\cdots j|j\cdots jj}{j}{\cdots}{j}\contraction{j\cdots j|j\cdots}{j}{}{j}\contraction{j\cdots}{j}{|}{j}\contraction{}{j}{\cdots j|j\cdots jjj\cdots j}{j}j\cdots j|j\cdots jjj\cdots jj\\
&=\frac{\delta(D_1)\cdots \delta(D_{n+m})}{L^{m+n}}+\cdots+\frac{\delta(\sum^{n+m}_{i=1}d_i)}{L^{2(n+m)-1}}\,.
\end{split}
\end{align}
In order to find the scaling in $L$, we need to sum over $x_i$ coming from the trace expressions, and determine how many deltas are redundant. We find
\begin{align}
\begin{split}
\sum^{L_A}_{x_1,\cdots,x_{2(n+m)}=1}\langle j(d_1)\cdots j(d_{2m})|j(d_{2m+1})\cdots j(d_{2(m+n)})\rangle\\
=\frac{{\cal O}(L_A^{m+n+1})}{L^{m+n}}+\cdots+\frac{{\cal O}(L_A^{2(m+n)})}{L^{2(m+n)-1}}\,,
\end{split}
\end{align}
where we only have a single redundant delta per term as in the previous case. Even though the two traces imply $\sum^{2m}_{i=1}d_i=0$ as well as $\sum^{2(m+n)}_{i=2m+1}d_i=0$, which should lead to some contractions with two redundant deltas, these terms are specifically excluded from the sum because they do not contain any crossing contractions. We therefore find that
\begin{align}
	f_{m,n}(L_A/L)=\lim_{L\to\infty}\frac{\langle\mathrm{Tr} [{\rm i}J]_A^{2m} |\mathrm{Tr} [{\rm i}J]_A^{2n} \rangle}{L_A}
\end{align}
is a polynomial function containing only powers $(L_A/L)^{m+n}$ up to $(L_A/L)^{2(m+n)-1}$. Moreover, in the limit $L\to\infty$, we have
\begin{align}
	\Sigma_S^2=\left(\frac{1}{\ln{2}}\right)^2\sum^\infty_{m,n=1}\frac{1}{L_A} \frac{f_{m,n}(L_A/L)}{4m(2m-1)4n(2n-1)}\,,
	\label{SigmaS2}
\end{align}
where $L_A^{-1} f_{m,n}(L_A/L)$ scales as $1/L$. Finding the coefficient analytically requires the summation of the full series. The coefficient can be computed numerically from a finite-size scaling analysis. 

In order to compute the variance scaling for a vanishing fraction $L_A/L\to 0$, it is sufficient to compute the first term in Eq.~(\ref{SigmaS2}). In fact, our analysis ensures that higher-order terms appear as higher powers in $L_A/L$. Truncating the series in Eq.~(\ref{SigmaS2}) at $m=n=1$ provides an expansion that is correct up to order $m+n-1=1$ in $L_A/L$. At this order, we find
\begin{align}
	\Sigma_S^2\sim \left(\frac{1}{\ln{2}}\right)^2\frac{1}{L}\left(\frac{L_A}{3L} +\frac{1}{6L_AL}\right)\,,
\end{align}
which is Eq.~(14) in the main text, evaluated in Fig.~2(b).

\section{Finite-size analysis of the Toeplitz Gaussian Ensemble}

\begin{figure}[!b]
\begin{center}
\includegraphics[width=0.99\columnwidth]{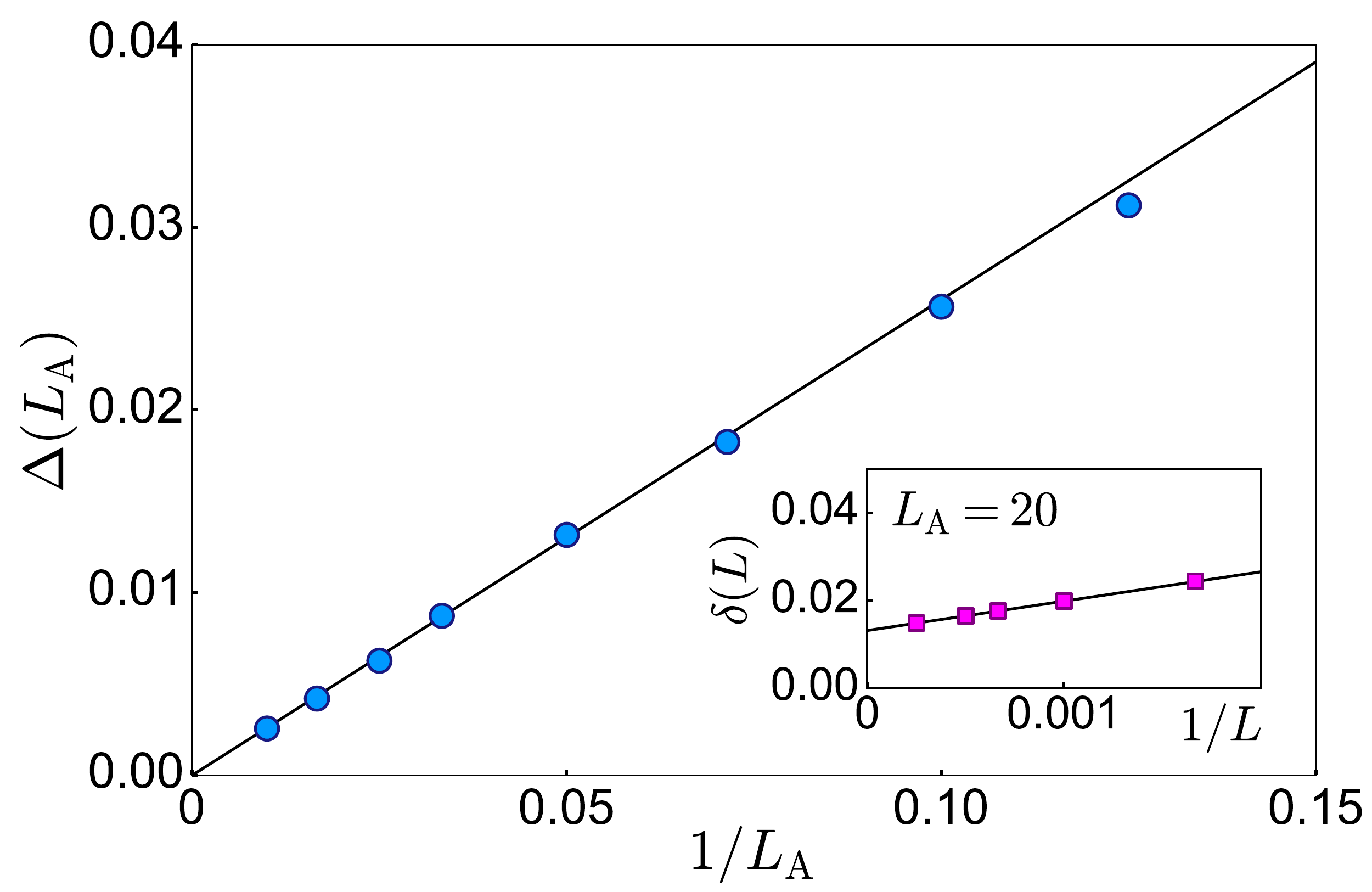}
\caption{ {\it Finite-size scaling of the difference of eigenvalue distributions between the Toeplitz Gaussian ensemble and the restricted complex structure eigenstates.}
We first extrapolate the difference $\delta(L)$ [see Eq.~(\ref{def_delta_L})] to $L\to\infty$ for a fixed value of $L_A$. This is shown in the inset for $L_A = 20$, where the solid line is a fit $\Delta(L_A = 20) + c/L$ to the results for $L \geq 1000$. [We get $\Delta(L_A = 20)=0.0132$ and $c = 6.691$.] Symbols in the main panel depict the values of $\Delta(L_A)$ as a function of $L_A^{-1}$. The solid line is a fit $0.260/L_A$ to the results for $L_A\geq 20$.
}
\label{figsup2}
\end{center}
\end{figure}

Figure~3 in the main text shows the distribution of eigenvalues ${\cal P}_\lambda$ of the restricted complex structure $[iJ]_A$. The numerical results for eigenstates are compared to the distribution of eigenvalues ${\cal P}_\lambda^{\rm (TGE)}$ of the Toeplitz Gaussian ensemble (TGE). The agreement between the curves is remarkable. Here we quantify their difference. We compute
\begin{equation} \label{def_delta_L}
 \delta(L) = \sum_\lambda \delta \lambda | {\cal P}_\lambda - {\cal P}^{\rm (TGE)}_\lambda |,
\end{equation}
where $\delta \lambda = 1/100$ is the width used when discretizing ${\cal P}_\lambda$.

We first observe that at fixed $L_A$, $\delta(L \to \infty)$ extrapolates to a small but finite value, see the inset of Fig.~\ref{figsup2} for $L_A = 20$. We use the fitting function $\delta(L;L_A) = \Delta(L_A) + c/L$ to obtain the coefficient $\Delta(L_A)$. In the second step, see the main panel of Fig.~\ref{figsup2}, we plot $\Delta(L_A)$ as a function of $L_A^{-1}$. The latter function extrapolates to zero when $L_A \to \infty$. In this limit, the eigenvalue distribution of the TGE therefore becomes identical to the eigenvalue distribution of the restricted complex structure $[{\rm i}J]_A$.

%\bibliographystyle{biblev1}
%\bibliography{references}

\newpage

\end{document}